\documentclass[a4paper, 11pt, twocolumn]{spie}
\usepackage[]{graphicx}
\usepackage{amsmath}
\usepackage{cite}

\title{Multiple field-of-view MCAO for a Large Solar Telescope: LOST simulations}
\author{Marco Stangalini, Francesco Berrilli, Dario Del Moro, Roberto Piazzesi
\skiplinehalf
Dipartimento di Fisica Universit\`{a} di Roma ``Tor Vergata'',\\
Via della Ricerca Scientifica, 1, Rome, Italy.\\
}
\authorinfo{Corresponding author: marco.stangalini@roma2.infn.it}
\begin{document} 
\maketitle 
\begin{abstract}
In the framework of a 4m class Solar Telescope we studied the performance of the MCAO using the LOST simulation package.
In particular, in this work we focus on two different methods to reduce the time delay error which is particularly critical in solar
adaptive optics:
a) the optimization of the wavefront reconstruction by reordering the modal base on the basis of the Mutual Information and 
b) the possibility of forecasting the wavefront correction through different approaches.
We evaluate these techniques underlining pros and cons of their usage in different control conditions by analyzing the results of the
simulations and make some preliminary tests on real data.
\end{abstract}
\section{INTRODUCTION}
In the next years the study of the Sun will be supported by the presence of a new class of large Solar telescopes (ATST, EST) that will shed new light over the small scale dynamics of the Sun. 
This class of telescopes will need efficient Adaptive Optics systems to achieve the corrected Field of View ($\sim 2'$) and high spatial resolution ($\sim 0.05"$) which are requested to chase after the major scientific targets.\\
In  particular, the absence of point-like sources on the solar surface slowes down the estimation of the wavefront and reduces the adaptive correction efficiency.
This is a very critical issue which has to be deeply analyzed in the Solar case.\\
In this work, we will focus on this point by showing a comparison between different approaches aimed to reduce the time lag in the MCAO correction loop.\\
Making use of both simulations and tests performed on real data sets, we study the behavior of different kinds of modal coefficients prediction tools based upon linear filters, neural networks and ARMA (Auto-Regressive Moving Average) processes.\\
We also show how the wavefront modal decomposition can be successfully optimized using information theory to speed up calculations and reduce the amount of information to process. 
\section{BASE OPTIMIZATION}
\label{sec:optim}
\subsection{Mutual Information Reordering} 
\label{subsec:info}
In the modal representation of the incoming wavefront it is necessary to both minimize the reconstruction error and to keep the number of modes in the representation basis low, allowing us to save computational time \cite{Fried1990}.
Therefore, the optimal basis is the one that can reproduce the wavefront with the smallest number of modes.
The Karhunen-Loeve (K-L) functions are the most popular choice, since they are ordered by covariance in a Principal Component Analysis (PCA) approach \cite{Dai, Dai1995} and in this sense usually perform much better than Zernike polynomials \cite{Noll76, beghi}.
We present an alternative tool to choose the best ordering for a modal basis which makes use of the concept of Mutual Information (MI).
MI has been used extensively in many applications involving relevant information extraction \cite{Baheti} or in constructing a metric for optimizing recognition tasks \cite{Baheti2}.
We have found that MI also allows us to reduce the basis dimension while maintaining the same error in the wavefront reconstruction with respect to covariance ordering \cite{stangalini}.\\
First, we introduce the MI concept: let us consider an information channel as a connection between source and receiver. The output $Y=\{y_{0},y_{1},y_{2},...,y_{J-1}\}$ of this channel (selected from alphabet $\widetilde{Y}$) can be thought of as a distorted version of input message ${X}=\{x_{0},x_{1},x_{2},...,x_{K-1}\}$, selected through alphabet $\widetilde{X}$.\\
The Shannon information entropy $H(\textit{X})$ \cite{Shannon}, is a measure of the a priori uncertainty about $X$, or even the average information content per source symbol:
\begin{center}
\begin{equation}
	H(\textit{X})=E[I(X)]=\sum_{k=0}^{K-1}p_{k}I(x_{k})=\sum_{k=0}^{K-1}p_{k}log(1/p_{k})
\end{equation}
\end{center}
where $p_{k}$ is the probability of occurrence of the event $x_{k}$, $I(\cdotp)$ is the information brought by the message and $E[\cdotp]$ is the expectation value.\\
We are interested in the measure of the uncertainty about the input $X$ after observing the output $Y$, that is we are interested in the estimate of the amount of information transferred through the channel.\\
To answer this question we can use the Mutual Information, defined as:
\begin{center}
\begin{equation}
	I(\textit{X};\textit{Y})= H(\textit{X})-H(\textit{X}|\textit{Y})
\end{equation}
\end{center}
where $H(\textit{X}|\textit{Y})$ is the conditional information entropy given by:
\begin{center}
\begin{equation}
	H(\textit{X}|\textit{Y})=\sum_{k=0}^{K-1} \sum_{j=0}^{J-1}p(x_{k},y_{j})log\left[\dfrac{1}{p(x_{k}|y_{j})} \right] 
\end{equation}
\end{center}
where $p(x_{k},y_{j})$ is the joint probability distribution function, and $x_{k}$ and $y_{j}$ are the input and the output respectively.\\
MI quantifies the reduction in the amount of uncertainty of the message $X$, knowing the output $Y$.
We can consider the phase reconstruction process as an information transfer from the unknown wavefront phase (source) to the polynomial basis.
In this framework the polynomial basis is the alphabet $\widetilde{Y}$ through which we can describe our wavefront phase, the reconstructed phase is a noisy version of the true phase and MI quantifies the amount of information transferred to the reconstructed phase.\\
We can therefore use MI to quantify the capability of every mode (Zernike polynomial or K-L function) to add information on the reconstruction of the incoming wavefront.
We are then able to reorder the modal basis according to the MI added by each mode so that the modes which carry the most MI come first.\\

\begin{figure*}[]
   \begin{center}
   \includegraphics[width=10cm]{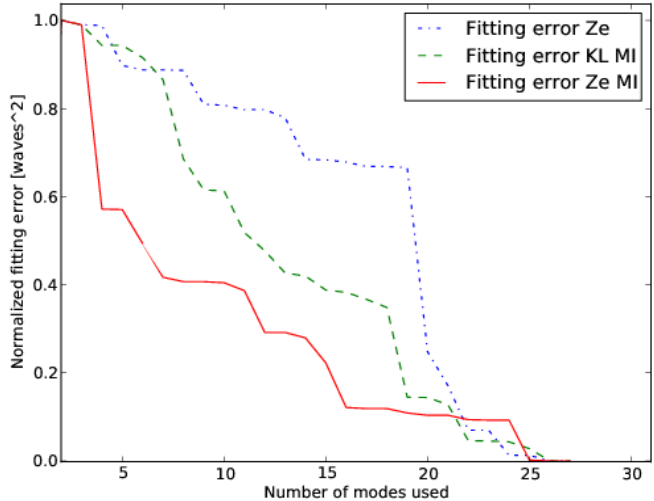} 
   \end{center}
   \caption{\footnotesize Wavefront fitting error vs number of modes used. Zernike polynomials standard ordering (dotted blue); K-L functions MI ordering (dashed green); Zernike polynomials MI ordering (continuos red). Results obtained from a tipical wavefront representation from the VTT dataset.}
\label{fit_error}
\end{figure*}

We have tested this method on Vacuum Tower Telescope \cite{VTT92, VTT05} closed loop wavefront sensor data . 
The dataset analyzed in this work consists of a time series of Zernike coefficients acquired on October 13th 2008, with the KAOS \cite{KAOS} wavefront sensor.
This wavefront sensor is characterized by 36 subapertures and allows us to perform a wavefront reconstruction with up to 27 Zernike modes.\\
We benchmarked the MI reordering against the standard ordering via a fitting error analisys, i.e. we estimated the residual errors of the phase reconstructed with an increasing number of modes. The residual error is defined as:
\begin{equation}
	\Delta \Phi(n)=|\Phi(n) - \Phi_{reference} |^2
\end{equation}
where $\Phi(n)$ is the wavefront reconstructed using the first $n$ modes ($1 \leq n \leq 27$) of the chosen basis and $\Phi_{reference}=\Phi(27)$ is the best representation of the real wavefront phase we have, i.e. the reconstruction with the full 27 Zernikes set.
Using this descriptor we tested three bases: the standard Zernike basis, a K-L basis reordered by MI and a Zernike basis reordered by MI.
The fitting error plots for a tipical wavefront representation from the dataset are presented in figure \ref{fit_error} as an example of the results.
In order to ease the interpretation of the plot, we recall that the steeper the fitting error function, the more performing the modal basis is, since it allows us to use less modes while maintaining the same error in the wavefront representation. In this example, we can see that a fitting error of $\simeq 0.4$ waves$^2$ can be obtained by reconstructing the wavefront with 20 Zernike polynomials (standard order), with 15 K-L functions (MI order) or with 7 Zernike polynomials (MI order).\\
MI reordering increases the performance for both Zernike and Karhunen-Loeve bases.
In particular, the Zernike MI reordered modal basis shows the steepest fitting error function.
This is somewhat unexpected, because it has been largely shown in literature that the standard K-L basis has a better performance than the Zernike one.
Our interpretation is that K-L modes bring high spatial frequency information, being defined by PCA as a linear combination of all Zernike modes, but, in closed loop situations, a deformable mirror reduces the phase variance, flattening the distortion amplitudes and thus suppressing the power at high spatial frequencies.
Therefore, in closed loop, the K-L representation may not be as efficient as in open loop applications.
%
\section{WAVEFRONT FORECASTING}
\label{sec:forec}
We have already introduced fact that the time delay reduces Adaptive Optics system performances; it is therefore extremely desirable to reduce the time delay by trying to forecast the next wavefront correction.
It has been demonstrated that WFS measurements are predictable \cite{Roddier:93, Jorgenson, Montera, Aitken1} and many open-loop demonstrations of the efficiency of such predictive methods have been given \cite{Jorgenson, Brockie}.
In the following we make use of three different approaches to try to forecast the time series of each Zernike coefficient.
\subsection{ARMA Approach} 
\label{subsec:ARMA}
In this section we introduce the ARMA (Auto Regressive Moving Average) predictive tools, which are a very common instrument in stationary time series forecasting.
For a deep treatment of this topic we suggest the reading of \cite{BoxJenkins:70}  and \cite{BrockwellDavis:87}.
Here we present only a brief introduction.\\
Our aim is to find $m(X_{n})$: a linear combination of the past $n$ values of the $X_{t}$ time series elements which is the ``best estimate'' of $X_{n+h}$.
We define as ``best estimate'' the function which minimizes the mean square error:
\begin{center}
      \begin{equation}
      	E|X_{n+h}-m(X_{n})|^{2}
      \end{equation}
\end{center}
where $X_{n+h}$ is the true value at the time $n+h$.
If we define ARMA($p$,$q$) processes as follows \cite{BrockwellDavis:87}:
\begin{center}
      \begin{equation}
      	X_{n}-\phi_{1}X_{n-1}-...-\phi_{p}X_{n-p}=Z_{n}+ \theta_{n1} Z_{n-1}+...+ \theta_{nq} Z_{n-q} 
      \end{equation}
\end{center}
where $Z_{t}$ is a white noise $0$ mean time series and $\phi_{t}$ and $\theta_{nt}$ represent the parameters of the autoregressive and moving average parts \cite{BrockwellDavis:87}, respectively, then it can be proved that the best prediction for $X_{n+1}$ can be written as:
\begin{center}
      \begin{eqnarray*}
      m(X_{n+1})=\phi_{1}X_{n}+...+\phi_{p}X_{n+1-p}+ \\
      + \sum_{j=1}^{n}\theta_{nj}\left(X_{n+1-j}-m(X_{n+1-j}) \right) \\
      if~ 1\leq n < m
      \label{prev2}
      \end{eqnarray*}
\end{center}
This implies that for an ARMA($p$,$q$) process, the computation of $m(X_{n+1})$ requires the knowledge of $k=max[p,q]$ previous values.
Since every process has a distinctive AutoCovariance Function (ACF), this can be used to characterize the process itself: we use the ACF to estimate the parameters of the process using the moments algorithm for the moving average parameters and the Yule-Walker algorithm for the autoregressive parameters \cite{chatfield}.\\
Our implementation of the ARMA tool can be summarized as follows:
\begin{itemize}
\item Find the best analytical ACF which fits the observed data ACF and estimates the ARMA parameters $\phi_{t}$ and $\theta_{nt}$ from that ACF as in \cite{BrockwellDavis2}.
\item Compute the forecast values as a linear combination of the past measurements with weights given by the ARMA parameters $\phi_{t}$ and $\theta_{nt}$.
\end{itemize}
The results of the application of this forecasting tool are shown in \ref{LOST}.
\subsection{Supervised Neural Network Approach}
\label{subsec:neural}
Another possible approach for forecasting is through Neural Networks (NN).
While with ARMA forecasting we have to estimate the process by fitting its ACF, neural networks are a more flexible paradigm which does not require any assumption on the process itself.
For an introduction to NNs we refer to the work of \cite{BrockwellDavis:87}.
Our NN forecasting procedure is implemented via a three layers supervised neural network using the sliding window technique and each measurement in the sliding window is addressed to one input neuron.
The network dimension in terms of learning speed, accuracy and stability has been chosen in order to maximize performance while keeping computational time as low as possible.
This resulted in a network with $10$ input neurons, $10$ hidden neurons and $10$ learning patterns.\\
In the backpropagation technique learning process \cite{Muller}, $k$ input terms $X_{n-k},..., X_{n}$ are fed to the network and the output $X_{n+1}$ is calculated using weights which are iteratively adjusted to minimize the output error since, in the learning stage, we exactly know the output.
We train the network with $10$ patterns, each composed of $k=10$ terms of the time series to define the neural weights, we then use the NN to compute the one-step forward prediction $X_{n+1}$ of the signal $X_{t}$.\\
As the learning process is made on the data itself, it would be necessary to recompute the network weights whenever there is a significant change in the characteristics of the time series, i.e. it is necessary to take into account the non-stationarity of the physical process.\\
The results of the application of this procedure are shown in \ref{LOST}.
\subsection{Linear Prediction Approach}
\label{subsec:linear}
As we have already recalled, in AO systems in closed loop operation what normally happens is that the wavefront state is measured at time $n$ and the correction is applied to the deformable mirror at time $n+1$, when the wavefront has, in principle, changed.
The simplest approach, when using forecasting methods to counter this time delay, is to use the last measured values to find the best linear estimation of the process; that is, estimate the linear slope from the last two measurements of the time series $X_{t}$: $X_{n+1}= aX_{n}+bX_{n-1}$, with $a$ and $b$ as weights of the linear relation.
In the case of a fixed time delay, $a=2$ and $b=-1$.
Such a trivial approach is only valid for very short time scales, but if a high loop stability is achieved in the AO correction, we do not expect high frequency variations on the signal.\\
We introduced this method and present the results in \ref{LOST} mainly for the purpose of comparing the performance of the different forecasting tools.
\subsection{LOST simulations}
\label{LOST}
\begin{figure*}[]
   \begin{center}
	   \begin{tabular}{cc}
		   \includegraphics[width=8cm]{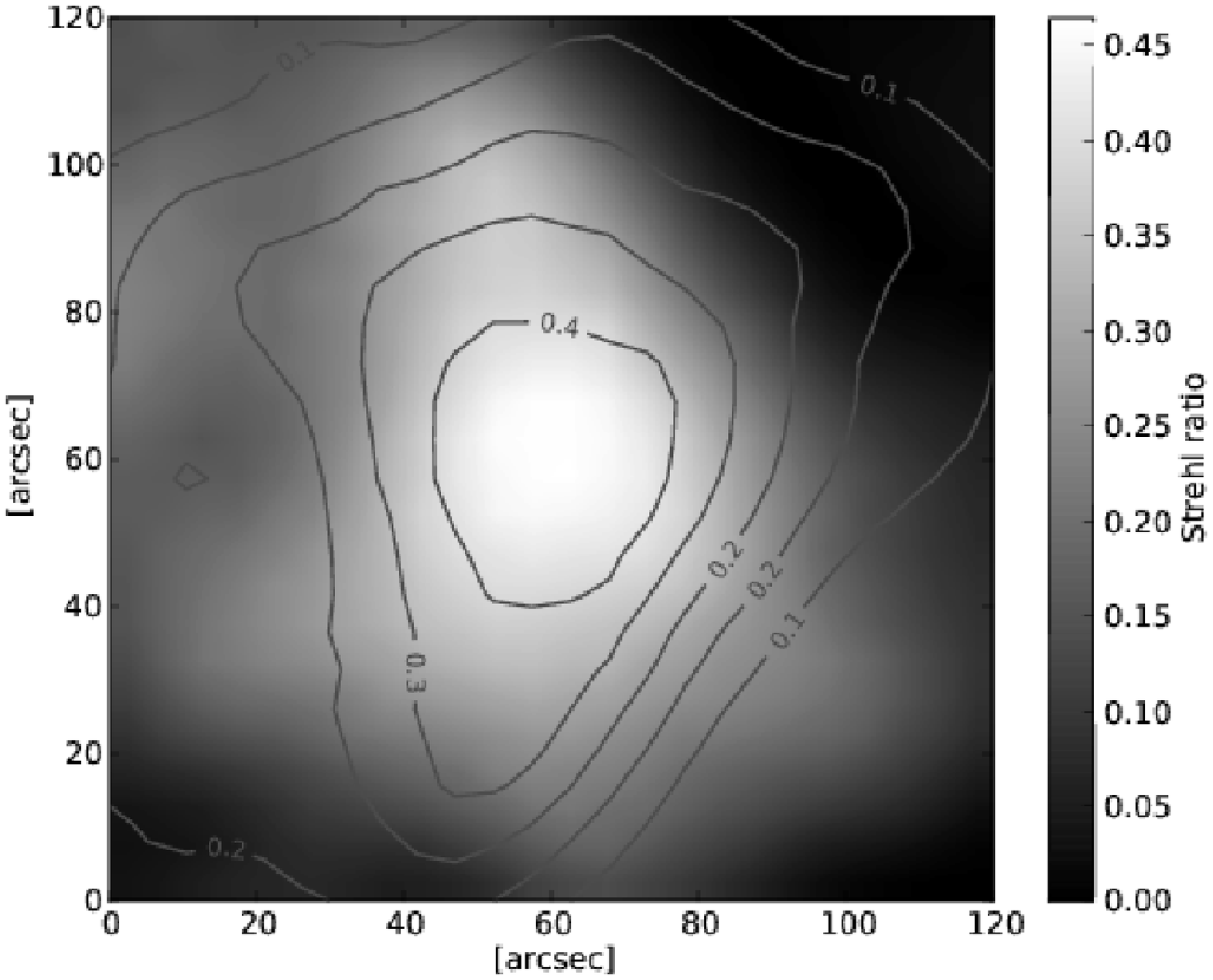} &  \includegraphics[width=8cm]{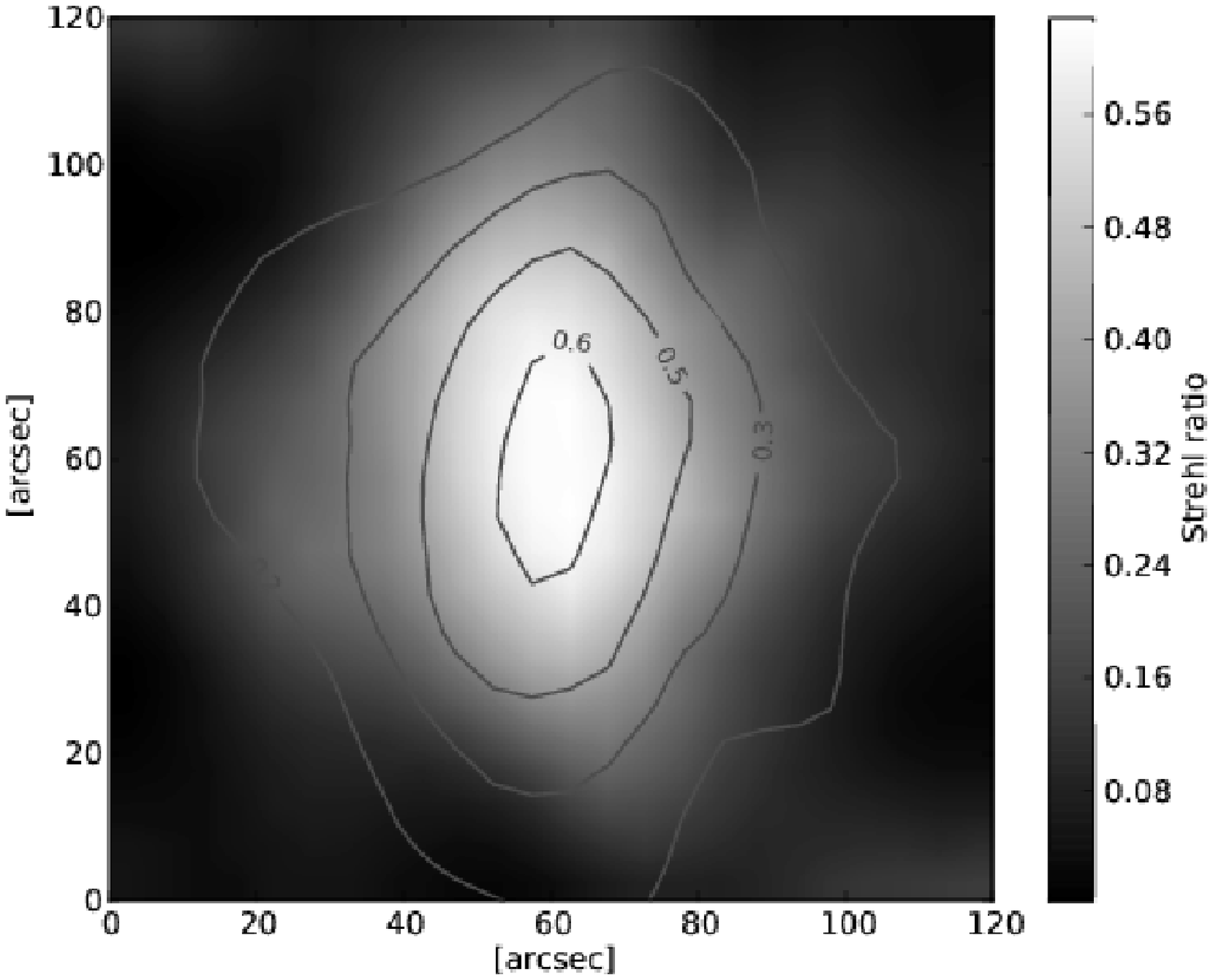}\\
		   \includegraphics[width=8cm]{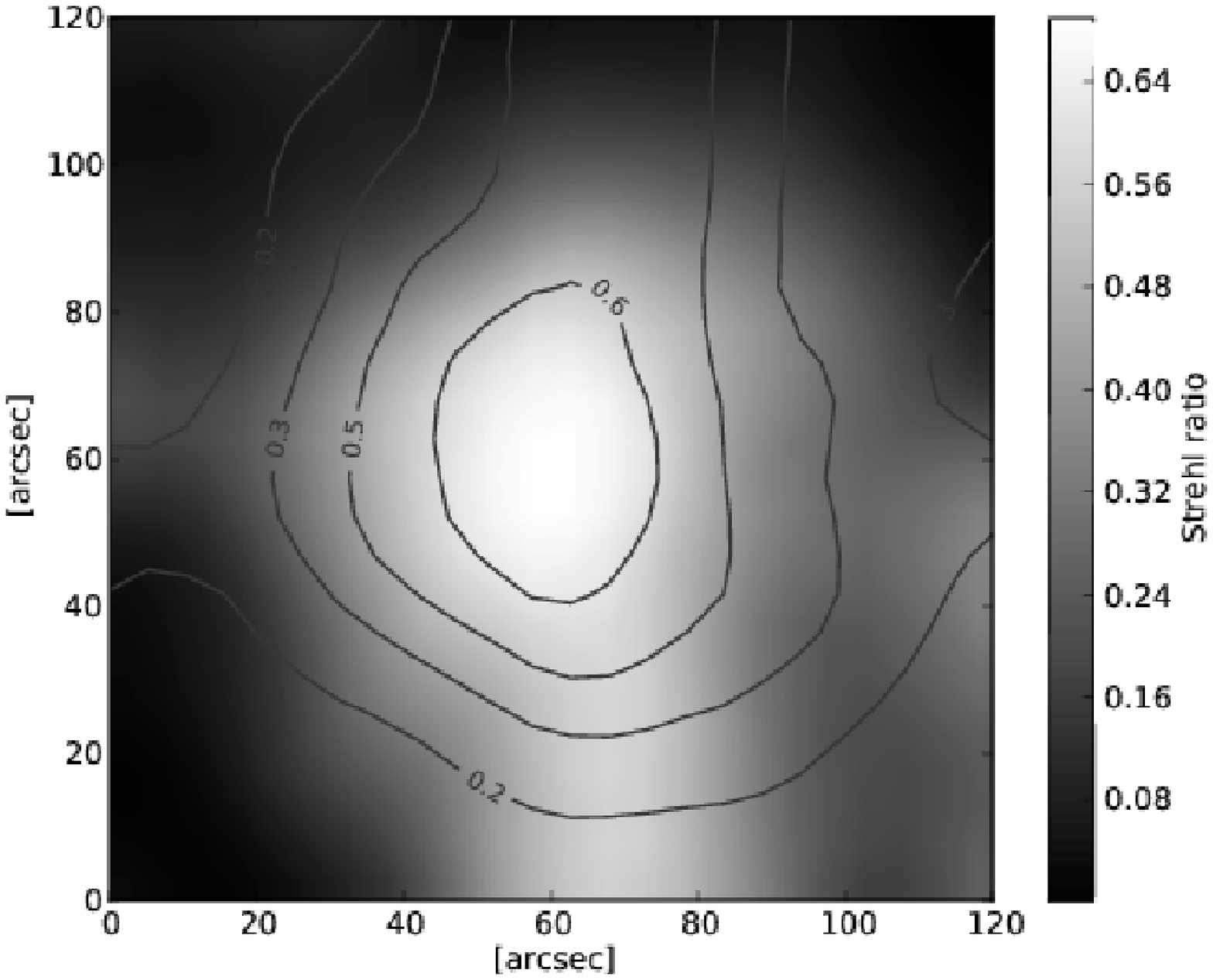}  &   \includegraphics[width=8cm]{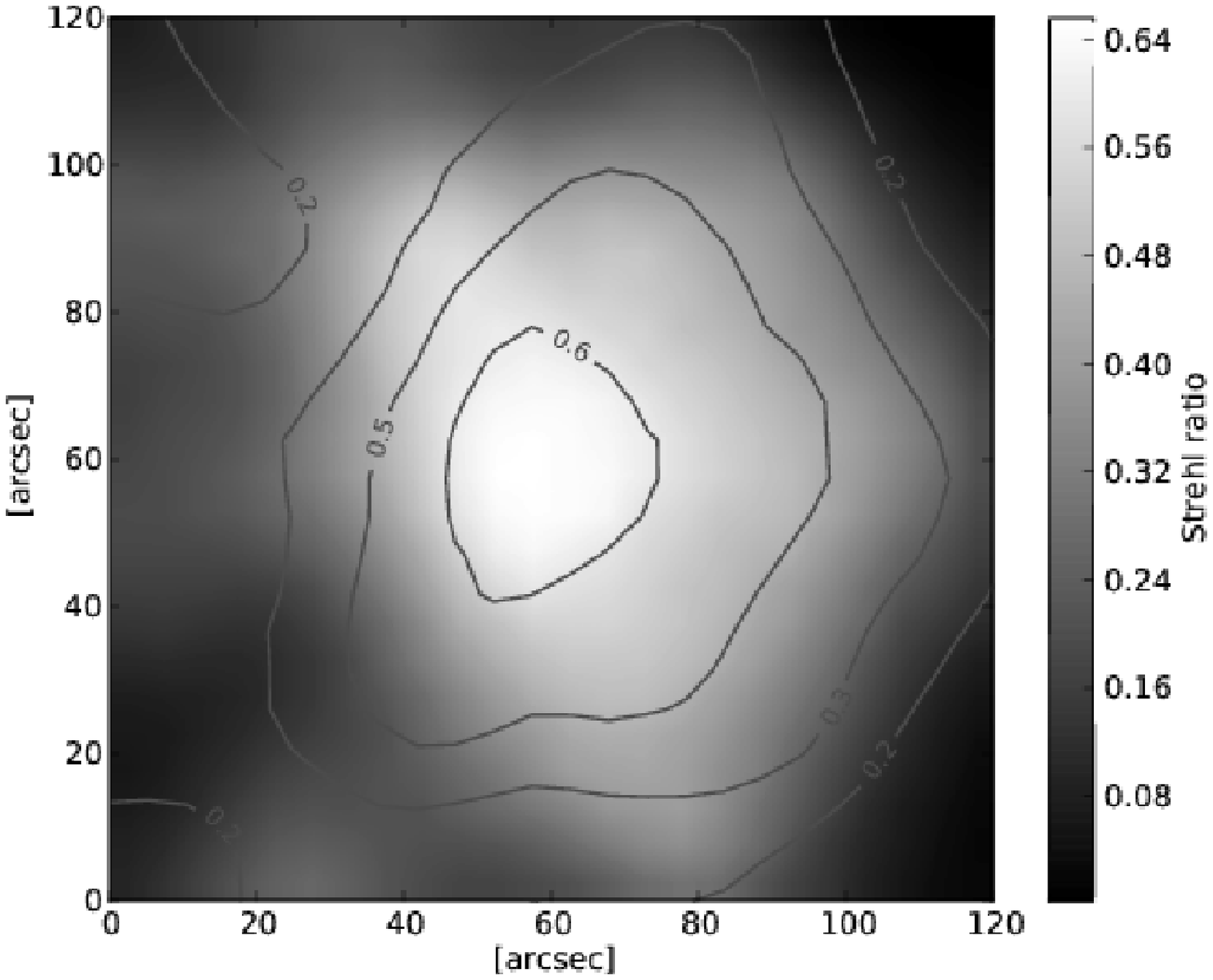}\\
	   \end{tabular}
   \end{center}   
   \caption{Average SR maps for the four simulation runs with SR isocontours overplotted. Upper left: standard LOST simulation; upper right: simulation with linear predictor; lower left: simulation with ARMA predictor; lower right: simulation with neural network predictor.}
	\label{SR}
\end{figure*}

To compare the behavior of these three prediction tools, we implemented them in the LOST (Layer Oriented Simulation Tool \cite{Arcidiacono2004}) simulation code to study how they affect the stability and the quality of the AO correction.
Given the LOST code capabilities of including the effects of Wave-Front Sensors (WFSs) on measurements, representing the phase delays introduced by the atmospheric layers in terms of phase screens and its modular design, it is a most convenient choiche to analyze the performance of a MultiConjugate Adaptive Optics (MCAO) system in a Multiple Field of View (MFoV) approach \cite{marchetti}.
The concept of MFoV MCAO was introduced in \cite{Ragazzoni2002}.
The idea is to use different WFS, each conjugated to different turbulence layers as in a layer-oriented approach, but with different FoVs for each sensor.
The WFS conjugated to the lowest turbulence layer (the ground layer) has the widest FoV, while the one conjugated with the highest layer has the narrowest.
Among other benefits, this approach fully exploits the photon flux on the WFS and allows a more uniform correction over the FoV.\\
The parameters of the LOST simulation we ran for the test are summarized in Table \ref{table1}.
Most of these parameters were chosen based on the main characteristics of current designs for future large solar telescopes (such as ATST and EST).
The parameters which describe the seeing have been set to reproduce a very simple atmosphere, which will give us nevertheless a good representation of the actual efficiency of the simulated MCAO.
In particular, in this simulation, there is perfect matching between the turbulent layers adding to the atmospheric distortions in the wavefront and the deformable mirrors compensating for them.
It is also worth to note that integration time and loop delay (which represent the amount of time allotted for the wavefront measure and for the application of the computed correction by the deformable mirror, respectively) together make up the total delay for the correction of the wavefront.
It is this total delay that the forecasting methods aim to counter.
\begin{table}[b]
\begin{center}
	\begin{tabular}{lc}
	\hline
	\hline
	 Telescope diameter & $4 ~m$ \\ 
	 Azimuthal angle & $0 ~\deg$ \\ 
	 Simulation time step & $1 ~ms$ \\ 
	 Number of deformable mirrors & $2$ \\ 
	 Integration time & $1 ~ms$ \\ 
	 Loop delay & $0.5 ~ms$ \\ 
	 Number of turbulence layers & $2$ \\ 
	 Type of turbulence & Kolmogorov \\ 
	 $D/r_{0}$ (lower layer) & $26.0$ \\ 
	 $D/r_{0}$ (upper layer) & $2.0$ \\ 
	 Wind speed (lower layer) & $7~m/s$ \\ 
	 Wind speed (upper layer) & $70~m/s$ \\
	\hline
	\hline
	\end{tabular} 
\end{center}
\caption{Main parameters of the LOST simulation}
\label{table1}
\end{table}
The prediction tools have been implemented in LOST as filters on the correction signals, so that they can compensate for the changes on the atmospheric turbulence conditions during the system actuation. 
It is essential to study their use in closed loop conditions because we expected them to change the correction signal properties dramatically, also from a statistical point of view.
\begin{figure*}[ht]
	\begin{minipage}[b]{0.5\linewidth}
		\centering
			\includegraphics[width=9cm]{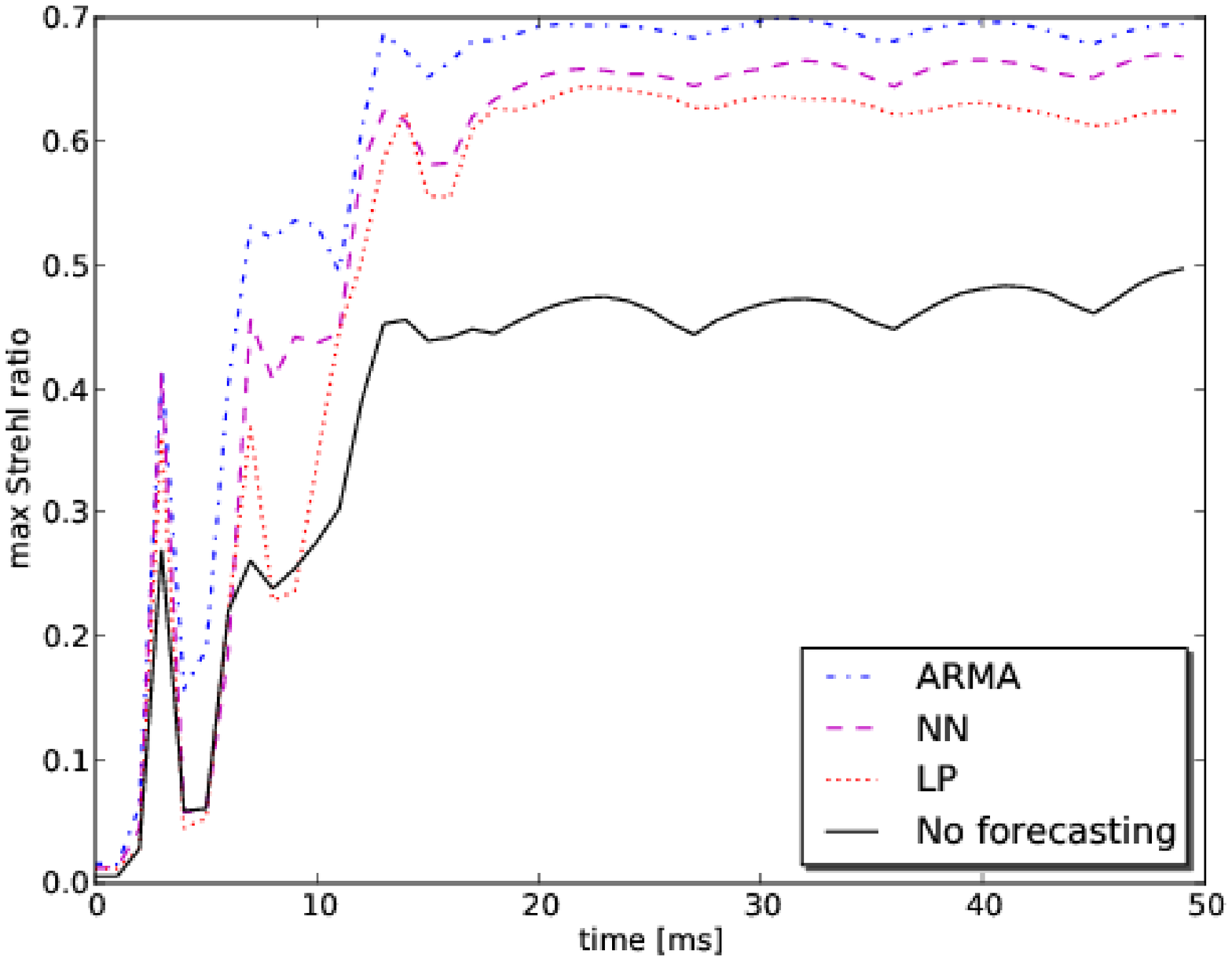}   
			\caption{Maximum SR vs time. Black continuous line: simulation without forecasting; red dots: linear predictor; magenta dashed line: neural network predictor; blue dash-dotted line: ARMA predictor.}
		\label{SRpeak}
	\end{minipage}
\hspace{0.5cm}
\begin{minipage}[b]{0.5\linewidth}
	\centering
		\includegraphics[width=9cm]{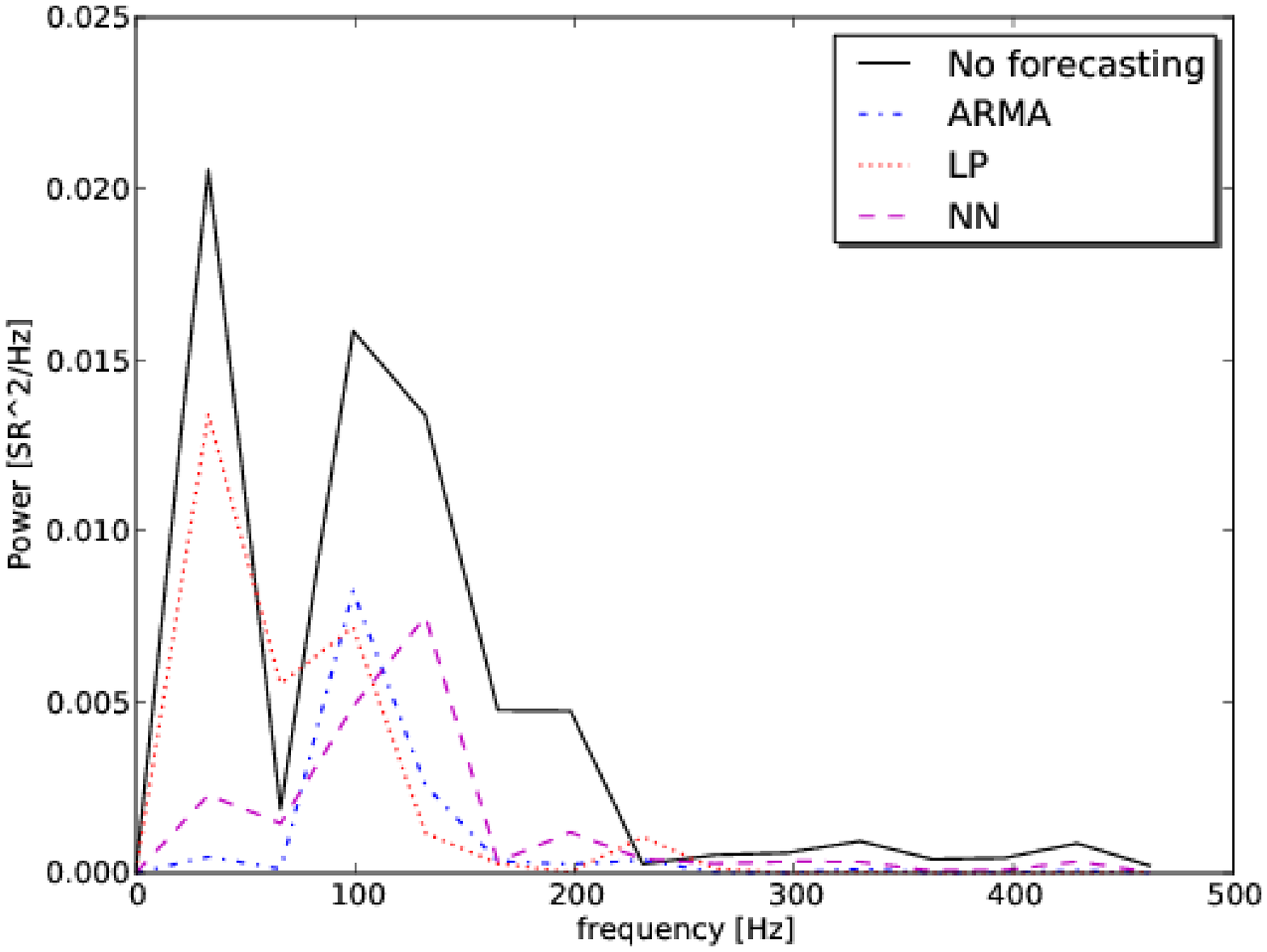}
		\caption{Power spectra of SR after the loop closure. Black continuous line: simulation without forecasting; red dots: linear predictor; magenta dashed line: neural network predictor; blue dash-dotted line: ARMA predictor.}
		\label{SRspectrum}
	\end{minipage}
\end{figure*}
Four LOST simulations were run with the same conditions, one for each of the forecasting tools and one with a standard MFoV MCAO correction.
We present here a study of both the evolution of the maximum Strehl Ratio (SR) value in the FoV and the fraction of the FoV in which the AO system is achieving a high correction efficiency.\\
From the standard MFoV MCAO simulation, we find that the ARMA model which fits the best the correction time-series in closed loop conditions is an ARMA(2,3) model. 
The $\phi_{t}$ and $\theta_{nt}$ parameter were estimated from the relevant ACF.
The standard simulation has been used also to train the NN: we used $10$ patterns of $10$ input values each taken after the AO loop closure ($time>15~ms$).
In Figure \ref{SRpeak} we show the behavior of the SR peak as a function of time for the four simulation runs.
After a few $ms$, the loop is firmly closed in all the four cases for the whole duration of the simulation.
The SR is strongly enhanced by all the prediction tools ($SR>0.6$) with respect to the performance without wavefront short-time prediction ($SR=0.4$), but the ARMA tool appears to perform best in terms of maximum SR achieved in closed loop conditions.\\
To deepen the analysis of the prediction tools on the loop stability, we studied the power spectrum of the oscillations of the SR peak after the loop closure (Figure \ref{SRspectrum}).
All three tools are able to reduce the power of the oscillations for frequencies above $70~Hz$ showing a similar behavior in the power spectra.
At low frequencies ($\sim 50~Hz$) only the ARMA and the NN approaches can damp out the SR oscillation power, while the linear prediction can only reduce the amplitude with respect to the standard correction.\\
As mentioned above, one of the most important requirements of the next generation Solar telescopes is a wide corrected FoV reaching $1'-2'$.\\
For this reason we also explored the effect of the prediction tools in expanding the corrected FoV.\\
In Figure \ref{SR}, we show the SR maps for the four different simulation runs averaged on the whole simulation run, but after the loop closure.
To help the reader, on the SR maps we overplotted SR isocontours.
As already stated, all the three methods can reach a higher SR peak, but it is straightforward to see how the prediction can greatly enlarge the corrected FoV.
Tipically, a $SR\simeq0.3$ is accounted as an acceptable value for AO performance, therefore, we refer to the area with a $SR>0.3$ as the FoV satisfactorily corrected by the MCAO.
We can estimate this area with an equivalent diameter: $d=\frac{4•\sqrt{A}}{\pi}$.
This diameter is extended from $d\simeq 0'.9$ in the standard MCAO to $d\simeq 1'.25$ by the linear predictor, to $d\simeq 1'.5$ by the NN predictor, to $d\simeq 1'.6$ by the ARMA predictor.\\
From the results of these simulations, it appears evident how the prediction tools could enhance the loop stability and achieve a better correction in terms of corrected FoV and SR peak.
In particular, the better performance of the NN and ARMA forecasting methods reflects their ability to describe the process underlying the wavefront time series using a more sofisticated approach.
\begin{figure*}[]
   \begin{center}
	   \begin{tabular}{ccc}
		   \includegraphics[width=5cm]{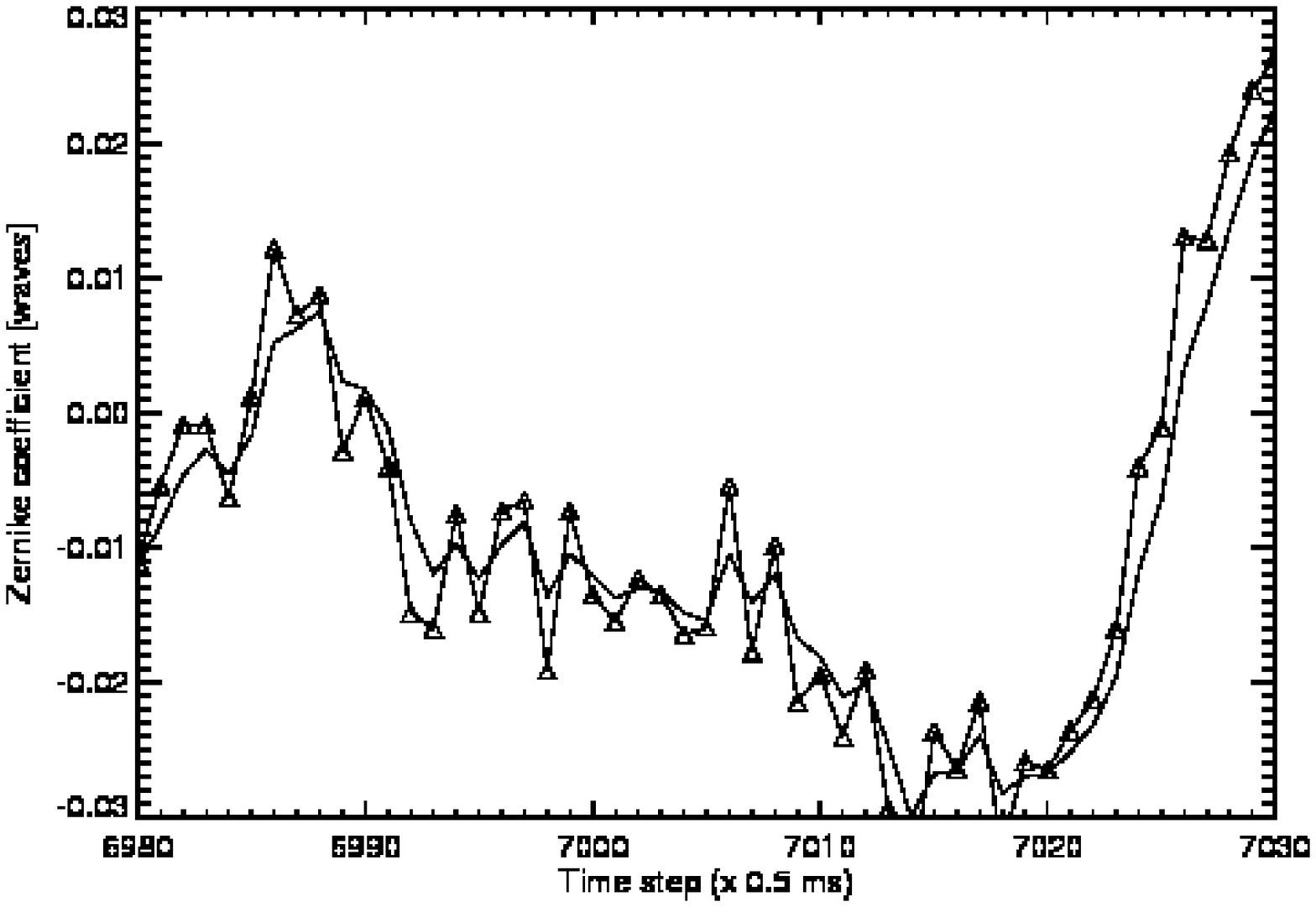} &  \includegraphics[width=5cm]{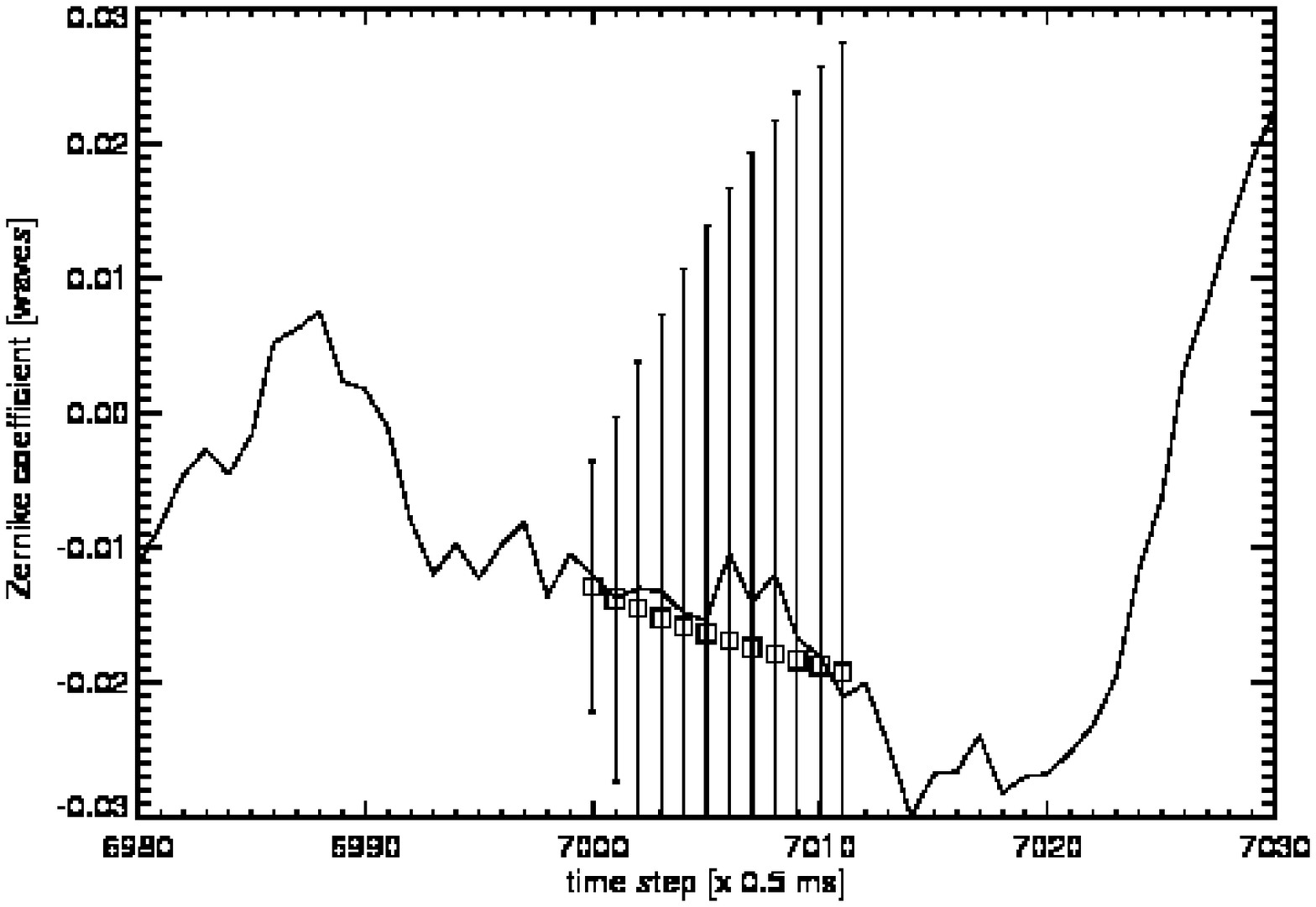} &  \includegraphics[width=5cm]{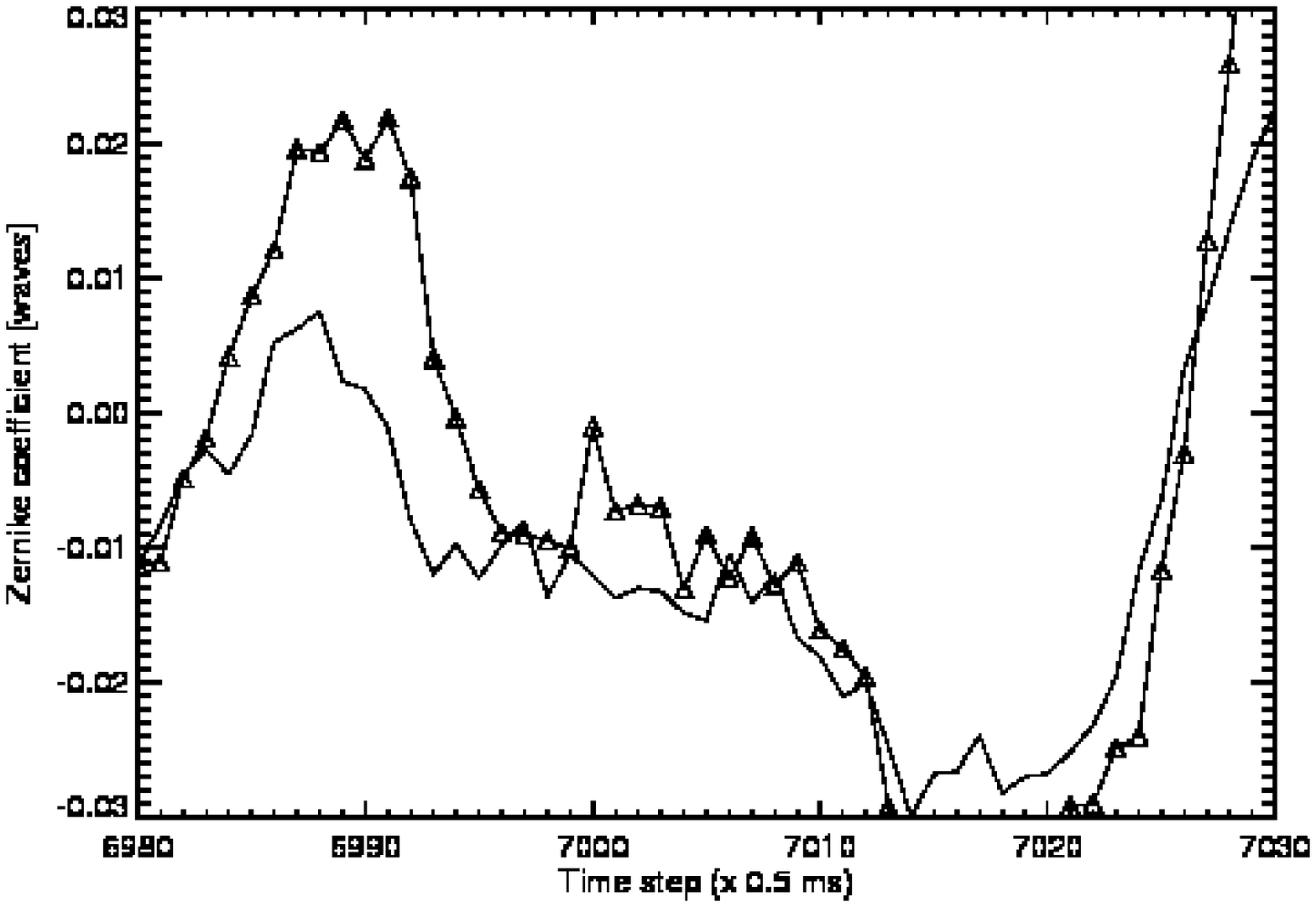}\\
	   \end{tabular}
   \end{center}
   \caption{VTT data. Left: Zernike 1 coefficient versus time with superimposed the LINEAR PREDICTOR prediction;
   						  center: Zernike 1 coefficient versus time with superimposed the ARMA prediction;
   						  right: Zernike 1 coefficient versus time with superimposed the NEURAL NETWORK prediction}
	\label{fore_fig}
\end{figure*}
\section{CONCLUSIONS AND FUTURE DEVELOPMENTS}
\label{subsec:forec_real}
In this work we have explored three methods to reduce the time delay in MCAO wavefront correction.
This time lag particularly affects the performances of those systems that cannot rely on high contrast or PSF-like sources and have to work with a poorly contrasted target as, for example, solar granulation. In these systems wavefront sensing requires the application of real-time correlation techniques which, in addition to the larger integration time needed, increase the time delay between measure and correction of the wavefront.\\
For these reasons, for the next generation solar telescopes like EST and ATST, methods are needed to reduce the time lag in order to achieve the desired MCAO performance.
We have shown, using MCAO simulations, that the use of prediction tools can dramatically enhance the MCAO performance in terms of corrected FoV, SR peak and loop stability.\\
A natural development of this approach would be the implementation of the prediction schemes by using dedicated hardware.
Another important step is of course the realization of an optical bench demonstrator to apply such methods in real conditions.\\
A first step toward this goal has been a preliminary test of the prediction tools on a real Zernike coefficient time series acquired with the KAOS wavefront sensor at the VTT Solar Telescope.
We present here an example of the performance of the three different prediction approaches in forecasting the same time series (Figure \ref{fore_fig}).
Here, we can not apply the forecast to the correction, therefore we have to verify whether the correction predicted on the basis of the measurements at or before time-step $n$ is a good representation of the measured Zernike coefficient value at time-step $n+1$.\\
The first plot refers to the linear predictor method, which shows a fairly accurate prediction of the signal, in fact the correction forecast for time-step $n+1$ is usually close to the signal at time-step $n+1$.\\
The second plot refers to the ARMA forecasting.
We have analyzed the Zernike coefficient time series and fitted its ACF in a time window of $3500~ms$, namely from time step $0$ to $7000$.
The model which best represented the time series was again an ARMA(2,3) model and it was used for $\phi_{t}$ and $\theta_{nt}$ parameter estimation.
In the plot we show how, using only time-steps $6998-7000$, ARMA is able to forecast the values from $7001$ to $7010$, with the associated error.
The approach shows a good prediction accuracy, at least on short time scales ($2-3~ms$) with an increasing error at longer time steps.\\
In the rightmost plot we show the forecast obtained by the NN described in \ref{subsec:neural}.
The learning time of the NN ended at time-step $6980$ and the triangles represent the prediction of the Zernike coefficient value $X_{n}$ computed using the previous $10$ values.\\
From preliminary results, it appears that the NN weights and ARMA parameters are valid only on time scales shorter than a few tens of $ms$ after the learning time.
Consequently, readjusting the NN weights and the ARMA parameters (or the whole ARMA model) periodically to account for the non-stationarity effects of the time series seems to be unavoidable.
Of course, in both cases, the ratio of the time spent between successive recomputations and the time required for the recomputation itself is the key.\\
Since the prediction tools act on the time series of coefficients used for the wavefront representation, it can be very useful to provide an efficient signal compression which can reduce the amount of information to process, while mantaining at the same time a high wavefront description accuracy.
Using information theory, we have also shown that, in the case of closed loop operation, there exists a more efficient representation than covariance ordererd K-L functions.
In particular, using Mutual Information, we are able to describe a turbulent wavefront with the same errors, but with a lower basis dimension by neglecting those terms which contribute the least to the wavefront description.\\
The next step is therefore to study the possible synergy of the two approaches: compressing the basis dimensionality and at the same time applying forecasting techniques to achieve the lowest possible delay in the application of the wavefront correction.
\bibliography{myrefs.bib}   
\bibliographystyle{spiebib}   
\end{document}